\documentclass[aps,prl,showpacs,twocolumn,amsmath,amssymb,floatfix, nobibnotes]{revtex4-2}

\usepackage{placeins} 
\usepackage{afterpage}
\usepackage{titlesec}
\usepackage[unicode=true, colorlinks=true, citecolor={blue!80!black}, urlcolor={blue!50!black}, linkcolor = {blue!80!black}]{hyperref}
\usepackage{amsmath}
\usepackage{amssymb}
\usepackage{graphicx}
\usepackage{bm}
\usepackage{dsfont}
\usepackage{xcolor}
\usepackage[utf8]{inputenc}
\usepackage[normalem]{ulem}
\usepackage{physics}

\newcounter{appendixpar}
\renewcommand{\theappendixpar}{Appendix~\Alph{appendixpar}}

\newcommand{\appendixpar}[1]{%
  \refstepcounter{appendixpar}%
  \textbf{\theappendixpar}: \textit{#1}.---%
}

\newcommand{\cpar}[1]{%
    \textit{#1}.---%
}

\newif\ifSMincluded
\SMincludedfalse

\SMincludedtrue

\newcommand{\SMcitekey}{SupplementalMaterial}  

\newcommand{\SMref}[2]{%
  \ifSMincluded
    \hyperref[#1]{#2}%
  \else
    #2~\cite{\SMcitekey}%
  \fi
}

\newcommand{\beq}{\begin{equation}}
\newcommand{\eeq}{\end{equation}}
\newcommand{\beqa}{\begin{eqnarray}}
\newcommand{\eeqa}{\end{eqnarray}}

\begin{document}

\title{Absence of Quasi-Majorana False Positives in Full-Shell Hybrid Nanowires}
\author{Carlos Payá}
\thanks{These authors contributed equally to this work.}
\affiliation{Instituto de Ciencia de Materiales de Madrid (ICMM), CSIC, 28049 Madrid, Spain}
\author{César Robles}
\thanks{These authors contributed equally to this work.}
\affiliation{Instituto de Ciencia de Materiales de Madrid (ICMM), CSIC, 28049 Madrid, Spain}
\author{Pablo San-Jose}
\affiliation{Instituto de Ciencia de Materiales de Madrid (ICMM), CSIC, 28049 Madrid, Spain}
\author{Elsa Prada}
\email{elsa.prada@csic.es}
\affiliation{Instituto de Ciencia de Materiales de Madrid (ICMM), CSIC, 28049 Madrid, Spain}

\begin{abstract}
Tunneling spectroscopy cannot be used as an unambiguous detection tool for Majorana zero modes (MZMs) in conventional partial-shell nanowires. The presence of smooth confinement at the end of the hybrid wire (among other sources of disorder) can create exponentially pinned zero-energy states, called quasi-MZMs, that mimic all local signatures of MZMs but lack topological protection. We find that this ambiguity in MZM detection does not occur in full-shell hybrid nanowires, an alternative nanowire design where a superconducting shell fully surrounds the semiconductor core. Acting as a synthetic vortex, a full-shell hybrid nanowire hosts Caroli-de Gennes-Matricon state analogs. In the presence of smooth confinement, these states create a topologically trivial skin at the wire's end that prevents the local probe from detecting quasi-MZMs. In this way, the same  mechanism that creates MZM impostors makes them invisible in tunneling spectroscopy, avoiding false positives produced by smooth disorder.
\end{abstract}

\maketitle

\cpar{Introduction} 
The search for Majorana zero modes (MZMs) in hybrid nanowires has been a subject of intense scrutiny by the condensed matter community due to their potential use as topologically protected parity qubits in future quantum computers \cite{Kitaev:P01, Nayak:RMP08, Beenakker:ARCMP13, Aguado:RNC17, Lutchyn:NRM18, Beenakker:SPLN20, Prada:NRP20, Marra:JoAP22, MicrosoftQuantum:PRB23}. In a pristine conventional hybrid nanowire, where a semiconductor core is partially covered by a $s$-wave superconductor and subjected to Zeeman field, a MZM is predicted to appear at each end of the wire when it enters the topological $p$-wave phase \cite{Lutchyn:PRL10, Oreg:PRL10}. The simplest way to detect these MZMs is through local spectroscopy \cite{Law:PRL09}. The appearance of a zero-energy peak (ZEP) in tunneling conductance \cite{Deng:NL12, Mourik:S12, Das:NP12, Deng:S16, Albrecht:N16, Gul:NN18} is considered a necessary, although not sufficient \cite{Bagrets:PRL12, Prada:PRB12, Kells:PRB12, Avila:CP19, Chen:PRL19, Prada:NRP20, DasSarma:PRB21, Hess:PRB21, Yu:NP21, Kouwenhoven:MPLB25}, condition for the existence of a Majorana bound state at the probed end of the wire.

Unfortunately, real hybrid nanowires are far from pristine. The presence of various types of disorder \cite{Pan:PRR20, Ahn:PRM21} to which $p$-wave superconductors are susceptible hinders the formation of well-behaved non-local MZMs located at the ends of the hybrid wire. Important sources of disorder include smoothly varying inhomogeneous potentials along the wire (created, e.g., by nearby gates) \cite{Kells:PRB12, Prada:PRB12, Rainis:PRB13, Roy:PRB13, Penaranda:PRB18, Fleckenstein:PRB18,  Vuik:SP19}, quantum dots \cite{Lee:NN14, Cayao:PRB15, Reeg:PRB17, Liu:PRB17, Rossi:PRB20, Hess:PRL23}, and strong, random electrostatic disorder (caused by chemical impurities, defects at the superconductor-semiconductor interface, etc.) \cite{Motrunich:PRB01, Bagrets:PRL12, Liu:PRL12, Pikulin:NJP12, Sau:PRB13, Mi:JETP14, Liu:PRB18, Lai:PRB19, Pan:PRB26}. Such types of disorder can give rise to zero-energy modes in the wire's spectrum (although of different nature and behavior) and thus to ZEPs in tunneling spectroscopy. These ZEPs are, in general, indistinguishable from those caused by true MZMs, but they have a trivial origin and thus lack topological protection. Therefore, the presence of these states prevents tunneling spectroscopy from providing conclusive evidence of MZMs, requiring more sophisticated experimental protocols \cite{Nichele:PRL17, Danon:PRL20, Pikulin:21, MicrosoftQuantum:PRB23,Sahu:PRB23}.

An alternative hybrid nanowire design capable of hosting MZMs is the full-shell nanowire, which has been under intense investigation in recent years \cite{Woods:PRB19, Vaitiekenas:PRB20, Vaitiekenas:S20,  Penaranda:PRR20, Kopasov:PSS20, Kopasov:PRB20, Sabonis:PRL20, Kringhoj:PRL21, Valentini:S21, Escribano:PRB22, Valentini:N22, San-Jose:PRB23, Paya:PRB24, Paya:PRB24a, Giavaras:PRB24, Vezzosi:SP25, Paya:PRB25, Deng:PRL25, Paya:PRB25a, Valentini:PRR25, Giavaras:PRB25}. It consists of a semiconductor nanowire fully coated by a superconductor shell and subjected to magnetic flux. In analogy to Abrikosov vortices in type-II superconductors, a full-shell nanowire hosts Andreev subgap states, dubbed Caroli-de Gennes-Matricon (CdGM) analogs \cite{Caroli:PL64, San-Jose:PRB23}. Full-shell nanowires offer some advantages over partial-shell ones, such as operating at much smaller magnetic fields \cite{Vaitiekenas:S20}, which helps preserve the superconducting state of the parent superconductor. This is because the mechanism driving the topological transition is the magnetic orbital effect \cite{Vaitiekenas:S20}, in contrast to partial-shell nanowires where it is driven by the Zeeman effect and requires much stronger fields.

While disorder in partial-shell nanowires has been extensively studied, its effects remain largely unexplored in full-shell geometries \cite{Vaitiekenas:S20, Valentini:S21, Valentini:N22, Paya:PRB24}. Here, we investigate the impact of an inhomogeneous electrostatic potential in Al/InAs full-shell nanowires. We show that a sufficiently smooth inhomogeneity induces quasi-Majorana zero modes (Q-MZMs) \cite{Kells:PRB12,Prada:PRB12,Vuik:SP19}, or partially-separated Andreev bound states \cite{Liu:PRB17}, analogous to those in partial-shells. The energy splitting of these modes is exponentially suppressed by the smoothness of the potential \cite{Kells:PRB12}, despite the two Q-MZMs strongly overlapping within the narrow region of electrostatic variation. Although the Q-MZM phenomenology in full-shells might intuitively be expected to parallel that of partial-shells, we reveal striking deviations between the two architectures.

In a tunneling spectroscopy setup with a full-shell nanowire, when tuning the system to a topological phase, a finite, topologically-trivial section appears between the insulating barrier and the bulk of the hybrid wire. Its width is proportional to the smoothness length. This region, which we dubbed a trivial \emph{skin}, is inherently connected to the presence of CdGM analogs and shifts the location of Q-MZMs to the wire's interior, making them invisible in practice to the local probe. Conversely, if the confining potential is not sufficiently smooth so that only a true MZM can form, the trivial skin reduces significantly and the MZM localizes closer to the wire's end. Consequently, the local tunnel probe becomes a faithful detector of MZMs. We analyze the physical mechanism for the appearance of the trivial skin in full-shell nanowires and explain why it is absent in partial-shell ones. 

\begin{figure}
   \centering
   \includegraphics[width=\columnwidth]{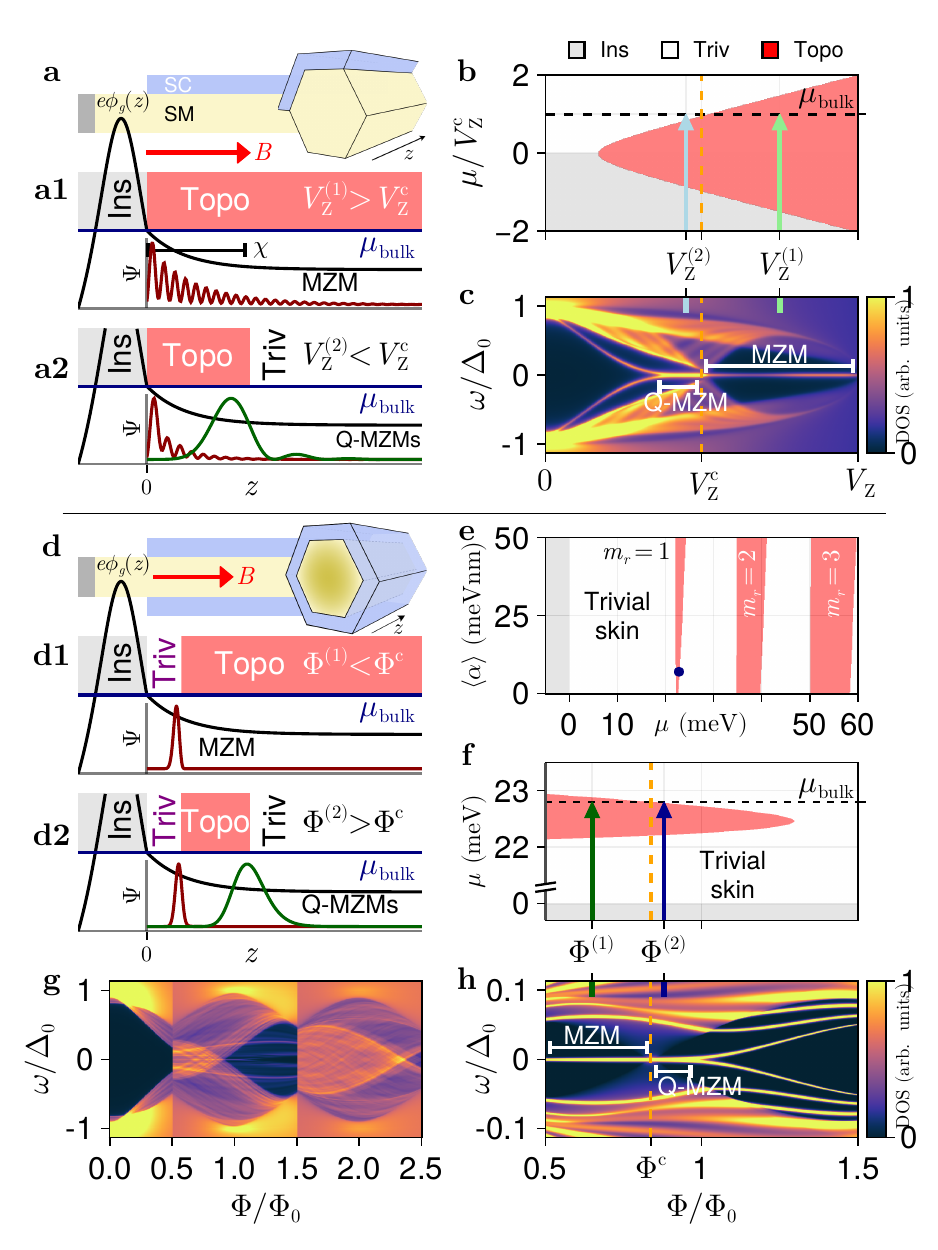}
   \caption{\textbf{Smooth electrostatic confinement in partial and full-shell nanowires.} (a) Schematic of a local spectroscopy setup for a semi-infinite partial-shell nanowire threaded by a magnetic field $B$ and with a gate-induced insulating tunnel barrier $\phi_g(z)$ that leaks into the nanowire over a length $\chi$. In panel (a1), given a chemical potential $\mu_\text{bulk}$, a Zeeman field $V_\text{Z}$ greater than a critical value $V_\text{Z}^c$ drives the entire wire into the topological regime, developing a  Majorana zero mode (MZM) at its end (red wave function). For a weaker field, in (a2), only a region of the wire becomes topological, developing quasi-Majorana zero modes (Q-MZMs) at its boundaries (red/green overlapping wave functions). (b) Topological phase diagram in the $(V_\text{Z},\mu)$ plane. (c) Density of states (DOS) versus $V_\text{Z}$ and energy $\omega$ (normalized to zero-field parent gap $\Delta_0$), highlighting MZM and Q-MZM zero energy peaks (ZEPs) at either side of $V_Z^c$. (d) Same as (a) but for a full-shell nanowire threaded by a magnetic flux $\Phi$. Now, a trivial-skin segment always appears between insulating and topological regions. (e) Topological phase diagram versus $\mu$ and average SOC $\langle \alpha \rangle$ for radial subbands $m_r$ at $\Phi=\Phi_0/2$. (f) Same as (b) but as a function of $\Phi$ for $\langle \alpha \rangle$ in the blue dot in (e). (g) DOS in the $(\Phi,\omega)$ plane, which shows a Little-Parks (LP) modulation with period $\Phi_0$, the superconducting flux quantum. (h) Contribution of the $m_J = 0$ angular momentum sector to the DOS within the first LP lobe, showing MZM and Q-MZM ZEPs at either side of the topological transition at $\Phi^\text{c}$. Parameters are chosen to show representative scenarios in Al/InAs hybrid nanowires, and are given in \ref{A:parameters}.}
   \label{Fig1}
\end{figure}

\cpar{Smooth confinement in partial-shell nanowires}
In a conventional hybrid nanowire, a semiconductor core with strong spin-orbit coupling (SOC) $\alpha$ is placed in contact to an $s$-wave parent superconductor that induces, by proximity effect, a superconducting pairing in the core \cite{Lutchyn:PRL10, Oreg:PRL10}. In modern hybrid nanowires, the superconductor is grown epitaxially as a thin shell over some facets of the semiconductor core, hence the name partial-shell nanowires. An applied axial magnetic field $B$ creates a Zeeman splitting $V_Z$ for the wire's electrons. When this field is larger than a certain critical value $V_Z^c$, the system undergoes a topological phase transition and a Majorana bound state appears at each wire's end. For a pristine one-dimensional (1D) nanowire model, we have $V_Z^c=\sqrt{\Delta_0^2+\mu_\text{bulk}^2}$, where $\Delta_0$ is the parent superconductor gap at $B=0$, and $\mu_\text{bulk}$ is the wire's chemical potential.

A Majorana bound state at one end of the hybrid nanowire can be probed using local spectroscopy, as illustrated schematically in Fig.~\ref{Fig1}(a). A normal probe is connected to the hybrid nanowire ($z>0$) through a tunnel barrier that is created with a local gate near the exposed nanowire section. The hybrid nanowire region is electrostatically screened by the superconductor. The gate creates an electrostatic potential $\phi_g(z)$ shown with a black line in Figs.~\ref{Fig1}(a1,a2), which makes the wire's local chemical potential $\mu(z)=\mu_{\rm{bulk}}-e\phi_g(z)$ negative inside the barrier, depleting it of carriers and making it insulating \footnote{The actual fully-depleted region is slightly smaller than the $\mu(z)<0$ barrier region, since carriers typically seep into it a small finite distance, making the insulator-trivial boundary a crossover.}. The spatial variation of $\phi_g(z>0)$ within the hybrid wire is characterized by the length scale $\chi$, which we call the smoothness parameter. Note that for $z\gg\chi$, $\mu(z)\rightarrow\mu_{\rm{bulk}}$. More details on the electrostatic potential can be found in \ref{A:phi(z)}. 

In Supplemental Material \SMref{A:partial-shell}{S-I}, we introduce the Hamiltonian of this system. To set the language and for subsequent comparison to the full-shell nanowire case, we summarize the main results here. The topological phase diagram is presented in Fig.~\ref{Fig1}(b), displaying topological (red), trivial (white), and insulating (gray) phases as a function of Zeeman field and chemical potential. For a Zeeman field $V_Z^{(1)}$ larger than $V_Z^c$ and for a $\mu_{\rm{bulk}}$ at the black dashed line, the hybrid nanowire is in the topological regime and a MZM appears bound to the left end, whose wave function is shown in Fig.~\ref{Fig1}(a1). It is highly oscillating with Fermi momentum and decreases exponentially towards the bulk with the Majorana coherence length \cite{Klinovaja:PRB12, Fleckenstein:PRB18}. In this case, the entire hybrid wire is in the topological regime, indicated by the red shaded region in Fig.~\ref{Fig1}(a1). Keeping the same $\mu_{\rm{bulk}}$, but reducing $V_Z^{(2)}$ below $V_Z^c$, results in the situation depicted in Fig.~\ref{Fig1}(a2). Now only a small region at the left end of the wire can be considered topological (red shaded), in the sense that locally $V_Z^{(2)}>\sqrt{\Delta_0^2+\mu(z)^2}$. The length of this region is of the order of $\chi$, and the rest is trivial (white shaded). Two states appear in the region where $\mu(z)$ varies, approximately bound to its ends. They are typically highly overlapping and, thus, not topologically protected, although they have a certain degree of wave-function non-locality \cite{Penaranda:PRB18}. These are quasi-Majorana bound states and are depicted in red and green in Fig.~\ref{Fig1}(a2). The red one is abruptly confined to the left end, whereas the profile of the green one is smoother and centered at the topological-trivial interface \cite{Fleckenstein:PRB18, Penaranda:PRB18} \footnote{The wave functions displayed in Fig.~\ref{Fig1} are not schematic, but rather numerically simulated for the parameters highlighted with green and blue marks in Figs. \ref{Fig1}(c,h), respectively. However, in the quasi-Majorana cases, the left and right wave functions are not depicted to scale; the left one (in red) should be much taller than the right one (in green).}. Note that in parameter space, $\mu(z)$ in Fig.~\ref{Fig1}(a1) follows a trajectory depicted by the green arrow in Fig.~\ref{Fig1}(b), from an insulating to a topological region. However, in Fig.~\ref{Fig1}(a2), $\mu(z)$ follows the blue-arrow trajectory, from an insulating to a topological and eventually to a trivial region.

Finally, the density of states (DOS) is presented in Fig.~\ref{Fig1}(c) \footnote{Note that this is not the LDOS at the end of the wire, but the DOS integrated over the region where $\mu(z)$ varies. Calculating the DOS allows us to observe features such as the topological bulk gap closing and reopening, and the presence of Q-MZM ZEPs in the full-shell case, which would be invisible in LDOS.}. At $B=0$ we observe a superconducting gap given by $\Delta_0$. As $V_Z$ increases, a state detaches from the continuum and approaches zero energy before the topological phase transition at $V_Z^c$ occurs. This approximation to zero energy is exponential with the smoothness parameter $\chi$, see Ref. \onlinecite{Kells:PRB12}. Thus, for sufficiently large $\chi$ ($\chi\gg\hbar\alpha\mu_\text{bulk}/V_Z\Delta_0$), a Q-MZM peak (corresponding to two Q-MZMs) develops and remains pinned close to zero energy for a sizable window of Zeeman fields. After the bulk gap closing and reopening at $V_Z^c$, clearly visible in the DOS, a true ZEP appears corresponding to a topological MZM. Note that no Majorana oscillations are present since we are considering a semi-infinite hybrid nanowire. The Majorana and quasi-Majorana wave functions of Figs. \ref{Fig1}(a1,a2) are taken at the green and blue marks in Fig.~\ref{Fig1}(c), respectively.

\cpar{Smooth confinement in full-shell nanowires}
Figures \ref{Fig1}(d-h) present the equivalent analysis for a full-shell hybrid nanowire. The model Hamiltonian and details of this geometry are provided in \SMref{A:full-shell}{S-II}. The tubular geometry of the superconductor and the accumulation of charge close to the superconductor-semiconductor interface introduce a number of differences with respect to the partial-shell geometry. The most important is the quantization of the fluxoid given by $n(\Phi) = \lfloor \Phi/\Phi_0\rceil$, where $\Phi_0=h/2e$ is the superconducting flux quantum. This integer represents the number of times the superconductor phase winds around the wire axis for a given flux. This quantization gives rise to the Little-Parks (LP) effect \cite{Little:PRL62, Parks:PR64}, whereby the shell gap oscillates with flux with periodicity $\Phi_0$ forming a series of so-called LP lobes labeled by $n=0,\pm1,\pm2...$, see Fig.~\ref{Fig1}(g). Within this gap, Andreev states coming from the different semiconductor transverse subbands typically populate the different lobes. These are CdGM analogs and have been extensively studied and  experimentally demonstrated in Refs. \cite{San-Jose:PRB23, Deng:PRL25}. In a cylindrical approximation, CdGM analogs are labeled by generalized angular momentum and radial quantum numbers $(m_J,m_r)$, see \SMref{A:full-shell}{S-II}. Crucially, only the $m_J=0$ subband can host MZMs \cite{Vaitiekenas:S20, Paya:PRB24}, restricting their emergence to the odd LP lobes where this generalized angular momentum is allowed.

In the partial-shell 1D model, the boundary of the topological region of Fig.~\ref{Fig1}(b) is analytical and given by $V_Z=V_Z^c$. For a given $\Delta_0$, it has a wedge shape and is independent of SOC as long as $\alpha\neq 0$. 
In the full-shell case, the topological phase arises within lobe $n=1$ for $\Phi_0/2\leq\Phi<\Phi^c$. The expression for the critical flux $\Phi^c$ is more complicated than for $V_Z^c$ and depends not only on $\mu$ and $\Phi$, but also on the average SOC $\langle\alpha\rangle$ \cite{Paya:PRB24}. In Fig.~\ref{Fig1}(e) we show the topological phase diagram as a function of $\mu$ and $\langle\alpha\rangle$ (for a fixed flux $\Phi=\Phi_0/2$). Different topological regions (in red) appear for different radial subbands \footnote{Note that a topological region for the lowest radial subband $m_r=0$ cannot be seen because it occurs at unrealistically large values of SOC.}. In Fig.~\ref{Fig1}(f) we show the topological phase diagram versus flux (within the $n=1$ LP lobe) and $\mu$ [for the SOC marked with a blue dot in Fig.~\ref{Fig1}(e)] \footnote{The topological region shown in Fig.~\ref{Fig1}(f) has a wedge shape, just like for the partial-shell nanowire, but oriented in the opposite magnetic field direction).}. Conspicuously, now there is a trivial (white) region between the insulating phase (gray) and the first accessible topological region (red) in both Figs. \ref{Fig1}(e,f). Thus, a finite chemical potential threshold $\mu^{ts}$ must be exceeded to access the topological phase, unlike the case presented in Fig.~\ref{Fig1}(b). We discuss this behavior extensively in \ref{A:skin}. In the physical full-shell hybrid nanwowire subjected to spatially varying $\phi_g(r,z)$, see \ref{A:phi(z)}, $\mu^{ts}$ creates a finite nanowire section between the insulating barrier and the topological region that is trivial in nature [in the sense that locally $\Phi<\Phi^c(z)$]. This trivial ``skin", in turn, pushes the Majorana and quasi-Majorana wave functions towards the interior of the hybrid wire, see Figs. \ref{Fig1}(d1,d2), respectively. 

The $m_J=0$ DOS for the blue-dot parameters of Fig.~\ref{Fig1}(e) is presented in Fig.~\ref{Fig1}(h). This is a blow-up around zero energy of the DOS in Fig.~\ref{Fig1}(g), but focusing on the $n=1$ LP lobe and considering only the $m_J=0$ angular sector. This DOS resembles the partial-shell DOS of Fig.~\ref{Fig1}(c), but there are some differences:
i) There is a bulk gap closing and reopening of the $m_r=1$ radial mode at a certain critical flux value $\Phi^c$ (marked with a vertical dashed line), in analogy with the partial-shell case, but now the MZM peak appears for magnetic fields smaller than this critical value, and the Q-MZM peak for larger ones. In other words, the Majorana--quasi-Majorana phenomenology is inverted with respect to the partial-shell case.
ii) There is actually no gap between the ZEPs and the continuum of states, as they coexist with a small DOS background coming from other filled radial modes. The background is further amplified when including all filled angular subbands $m_J$ as in Fig.~\ref{Fig1}(g).
%ii) When the total DOS is considered [i.e., including all filled angular subbands $m_J$ like in Fig.~\ref{Fig1}(g)], there is typically no minigap between the ZEPs and the continuum of states. These ZEPs coexist with a finite DOS background coming from the CdGM analogs.

\begin{figure}
   \centering
   \includegraphics[width=\columnwidth]{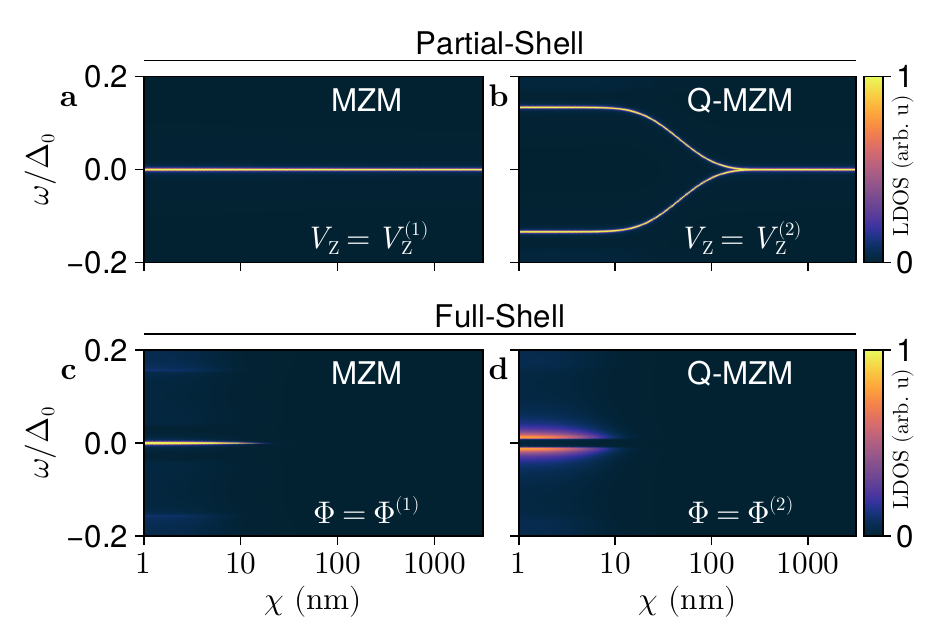}
   \caption{\textbf{Detection of MZMs versus Q-MZMs.}
   LDOS of the $m_J = 0$ sector versus smoothness parameter $\chi$ and energy $\omega$ at the end of a semi-infinite partial-shell (a,b) and full-shell (c,d) hybrid nanowires. The left (right) column corresponds to the Majorana (quasi-Majorana) case. In (d), the visibility of low-energy states is lost before the Q-MZM peak if formed. Parameters taken at the green and blue marks of Fig.~\ref{Fig1}(c,h), respectively. Other parameters in \ref{A:parameters}.
   }
   \label{Fig2}
\end{figure}

\cpar{Local density of states of topological MZM and trivial quasi-MZMs}
In the tunneling regime, the differential conductance through the normal-insulator-superconductor system is proportional to the LDOS at the end of the hybrid wire \cite{Datta:95}, and thus one can access the local spectrum at $z=0$. 

In Fig.~\ref{Fig2}(a) we show the LDOS at the end of a semi-infinite partial-shell hybrid nanowire for the Majorana case $V_Z=V_Z^{(1)}$ of Fig.~\ref{Fig1}(a1). The LDOS is plotted against energy and smoothness parameter $\chi$ (given on logarithmic scale). A robust ZEP is consistently found for every $\chi$, as corresponds to a topological state. The spatial variation of the electrostatic potential affects the profile of the Majorana wave function, but it is always clearly detected in LDOS when the whole wire is in the topological regime. On the other hand, for $V_Z=V_Z^{(2)}$ in Fig.~\ref{Fig1}(a2), a zero-energy Q-MZM only appears when $\chi$ is sufficiently large (otherwise, the two electronic states split in energy and thus cannot be mistaken for MZMs). However, when present, this ZEP is indistinguishable from the signature of a true MZM, rendering this measurement alone insufficient for definitive Majorana detection.

Applying this same analysis to a full-shell hybrid nanowire yields a striking result. For small values of $\chi$, a ZEP develops in the case of a true MZM, Fig.  \ref{Fig2}(c), while there is a split peak in the quasi-Majorana case, Fig.  \ref{Fig2}(d), just like in Figs. \ref{Fig2}(a,b), respectively. However, as the smoothness increases, the Q-MZM splitting is suppressed, but the visibility of the corresponding LDOS signals disappears altogether. There are no traces of low-energy modes. The reason is the formation of a sizable trivial skin between the contact and the topological region, which buries any MZMs or Q-MZMs and thus reduces their visibility at the end of the wire. Hence, false positives from Q-MZMs are avoided, and only MZMs at the end of topological and sharply confined  nanowires will be detectable (buried topological MZMs will remain hidden). These conclusions are confirmed by tunneling differential conductance simulations (at various temperatures) where we include the contribution of all $m_J$ modes, see \SMref{A:conductance}{S-IV} and \SMref{A:temperature}{S-V}.

\begin{figure}
   \centering
   \includegraphics[width=\columnwidth]{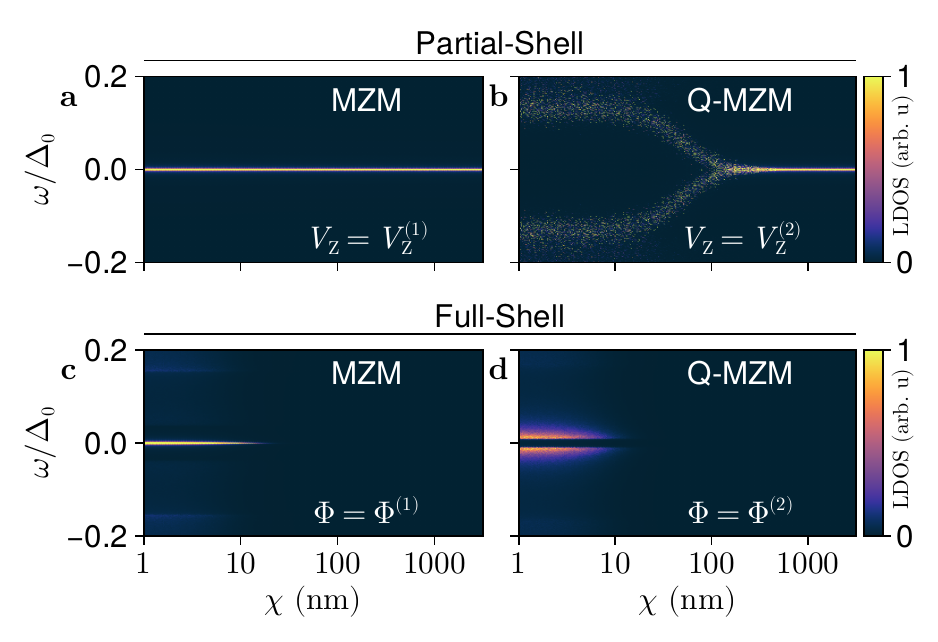}
   \caption{Same as Fig.~\ref{Fig2} but in the presence of short-range disorder. Parameters are given in \ref{A:parameters}.}
   \label{Fig3}
\end{figure}

So far, we have considered that the hybrid wire has a smooth potential inhomogeneity close to the probe, but is otherwise free from other sources of disorder. It has been argued that current experimental nanowires are plagued with impurities. The inclusion of short-range disorder smaller than the topological minigap, see \ref{A:phi(z)} and \ref{A:parameters}, yields the results shown in Fig.~\ref{Fig3}. ZEPs associated with true MZMs are resilient to this amount of disorder in both partial- and full-shell nanowires [Figs. \ref{Fig3}(a) and (c), respectively]. Non-topological peaks, however, become strongly smeared. Notice, e.g., that the Q-MZM in Fig.~\ref{Fig3}(b) disappears. Most importantly for our analysis, the absence of Q-MZM signals in local spectroscopy for the full-shell nanowire persists, Fig.~\ref{Fig3}(d).

\cpar{Conclusions}
Full-shell hybrid nanowires, consisting of a tubular superconducting shell surrounding a semiconductor core, behave like a synthetic vortex. As conventional vortices, they enclose Andreev subgap states, which in this case are CdGM analogs. This means that, when a pristine hybrid nanowire is brought to a topological superconducting state by means of a magnetic flux, MZMs appearing at the wire's ends generally coexist with such Andreev states. In the presence of realistic sources of disorder, impostor Majorana states may appear. One of the most problematic is the Q-MZM that arises from sufficiently smooth electrostatic potentials, since it can remain exponentially pinned to zero energy within a finite region of parameter space, like true MZMs. However, CdGM analogs in full-shell nanowires with smooth potentials inevitably create a  topologically trivial skin at the wire's end (see \ref{A:skin}) that prevents the local probe from detecting Q-MZMs in practice. This does not happen in partial-shell devices. In conclusion, CdGM analogs guarantee the absence of Q-MZM false positives when interpreting pinned ZEPs in the tunneling spectroscopy of full-shell hybrid nanowires.

\cpar{Acknowledgments} We thank C. M. Marcus for insightful comments. This research was supported by Grants PID2021-125343NB-I00, PRE2022-101362, PID2023-150224NB-I00 and CEX2024-001445-S, funded by MICIU/AEI/10.13039/501100011033, ``ERDF A way of making Europe'' and ``ESF+''; and JAE program, funded by CSIC/10.13039/501100003339.

\cpar{Data availability} Data supporting the findings of this article are openly available. The code is based on \texttt{Quantica.jl} \cite{San-Jose:25a}. The specific code to build the nanowire Hamiltonian and perform and plot the calculations is available in Refs. \cite{Paya:26a} and \cite{Paya:26}, respectively. Some details are provided in \SMref{A:numerics}{S-III}. Visualizations were made with the \texttt{Makie.jl} package \cite{Danisch:JOSS21}. 

\nocite{Skalski:PR64,Schwiete:PRB10,Lutchyn:PRL11,Mikkelsen:PRX18,Antipov:PRX18}
\bibliography{Quasi-Majoranas,supp}

\pagebreak 

\onecolumngrid
\begin{center}
    \textbf{\large End Matter}
\end{center}
\twocolumngrid

%------------------------------------------------------------
% APPENDIX: Smooth confinement
%------------------------------------------------------------

\appendixpar{Smooth confinement model}
\label{A:phi(z)} 
The spatial profile of the electrostatic potential decaying from the tunnel barrier into the full-shell nanowire (running along the $z$ axis) depends on both the boundary conditions from gates and the internal charge distribution $\rho$. A rigorous treatment of this geometry requires a self-consistent Schrödinger-Poisson calculation to compute $\rho$ as a function of gates and geometry, but this is beyond the scope of the present work. Instead, we approximate the $z\geq 0$ electrostatic potential inside the hybrid wire (with a semi-infinite shell) as $\phi_\text{tot}(\vec{r})=\phi_\infty(r) + \phi_g(\vec{r})$, where $\vec{r}=(r,\varphi,z)$ in polar coordinates. Here  $\phi_\infty(r) = U(r)/e$ is a $z$-independent (and $\varphi$-independent) potential with a dome-like profile, treated phenomenologically and described in \SMref{A:full-shell}{S-II}. It accounts for the $\rho(r)$ accumulation towards the core/shell interface, such that $\nabla^2 \phi_\infty = -\rho/\epsilon$, where $\epsilon$ is the dielectric constant. The correction $\phi_g$ decays deep inside the nanowire and stems from the gates outside that deplete the semiconductor core of carriers in the tunnel barrier. Inside the hybrid wire, $\phi_g$ satisfies a Laplace equation $\nabla^2 \phi_g = 0$, since $\rho$ is already accounted for by $U$. This equation is supplemented by Dirichlet boundary conditions $\phi_g(R,z)=0$ all along the core/shell boundary, since the shell is assumed to be grounded. In addition, the tunnel barrier at $z<0$ imposes $e\phi_g(r, z =0) = \mu_\text{bulk}$ at the left end of the hybrid wire, where $\mu_\text{bulk}=U(R)$ is the asymptotic chemical potential deep inside the wire. This makes $\mu(r, z<0)$ negative, making the carrier density quickly decay to depletion inside the tunnel barrier.

Under these assumptions, the solution to Laplace's equation inside the cylindrical nanowire ($z>0$) is a sum of evanescent contributions with decreasing decay lengths,

\begin{equation}
    e\phi_g(r, z) = \mu_\text{bulk} \sum_{\nu = 1}^\infty \frac{2}{k_\nu J_1(k_\nu)}J_0\left(k_\nu \frac{r}{R}\right)e^{-\frac{k_\nu}{k_1} \frac{z}{\chi}},
\end{equation}
where $J_i$ is the $i$-th order Bessel function of the first kind, $k_\nu$ is the $\nu$-th zero of $J_0$ such that $J_0(k_\nu) = 0$, and $\chi = R/k_1$ is the decay length of the slowest exponential $\nu = 1$.

In the above treatment we have neglected one important aspect, the $z$ dependence of $\rho$ resulting from the screening of the $\phi_g$ potential itself. Instead of attempting a self-consistent treatment of this effect with a full Poisson solver, we model it phenomenologically by making the fundamental decay length $\chi$ a free parameter, dubbed smoothness parameter. The original value $\chi = R/k_1$ then corresponds to the case without screening along $z$. 

For a partial-shell nanowire, $\phi_g(\vec{r})$ takes a different form that strongly depends on the specific backgate potential along the wire, lacking the simple radial symmetry of the full shell. However, the generic exponential decay along $z>0$ persists and is expected to hold even in a full Schrödinger-Poisson treatment, albeit with corrected values. Since our primary focus in the main text is describing the effect of smooth longitudinal confinement on MZMs and Q-MZMs, for the partial-shell model we ignore the details of the cross-sectional variation of both $\phi_g$ and $U$ and retain only the fundamental contribution to $\phi_\text{tot}(\vec{r})=\phi_g(z)$ with the longest decay length, $\nu = 1$. The confinement potential energy then simplifies to:
\begin{equation}
    \label{eq:UPS}
    e\phi_g(z) = \mu_\text{bulk} e^{-z/\chi},
\end{equation}
where once more $\chi$ is a phenomenological smoothness parameter or screening length. 

In the last part of the main text (Fig.~\ref{Fig3}), apart from a smooth confinement, we consider the additional presence of strong, short-range, electrostatic disorder. The potential energy is then $e\phi_\text{tot}(\vec{r})+U_{\rm{imp}}(z)$, where $U_{\rm{imp}}(z)$ is modeled as an uncorrelated random variable with a uniform distribution of amplitude $A_{\rm{imp}}$ and zero mean value. For numerical efficiency, $U_{\rm{imp}}(z)$ is present only in the region where the confinement potential varies, and is truncated once the longitudinal potential reaches its bulk value.

%------------------------------------------------------------
% APPENDIX: Trivial skin
%------------------------------------------------------------

\appendixpar{Origin of the trivial skin}
\label{A:skin} 
The full-shell nanowire Hamiltonian is diagonalized in generalized angular momentum modes, $m_J$, which take integer values for \emph{odd} $n$ and half-integer values for \emph{even} $n$ (see \SMref{A:full-shell}{S-II}). The relevant topological mechanism for these wires is driven by magnetic flux instead of the Zeeman effect and, crucially, requires $m_J = 0$ \cite{Paya:PRB24}. Therefore, only this specific mode within the \emph{odd} LP lobes can undergo a topological transition. If other $m_J \neq 0$ subbands become occupied before the $m_J = 0$ mode as the carrier density is increased, the proximity effect from the shell will make these subbands a trivial superconductor (possibly gapless). As the chemical potential $\mu$ is increased, starting from a $\mu=0$ depleted nanowire, the topological regime then requires exceeding a finite value, $\mu^\text{ts}$, measured from the band bottom.

\begin{figure}
   \centering
   \includegraphics[width=\columnwidth]{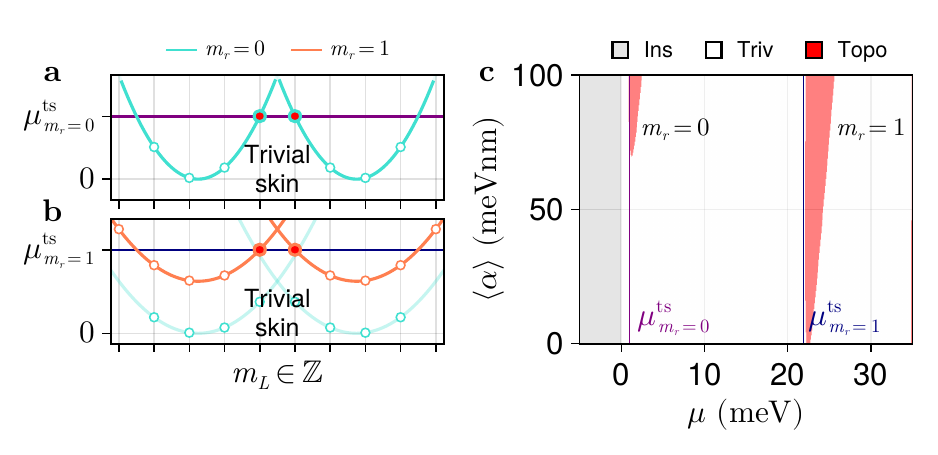}
   \caption{\textbf{Origin of the trivial skin.} (a) Schematic representation of the topological filling of the first radial subband ($m_r = 0$) of a full-shell nanowire as a function of angular momentum $m_L \in \mathbb{Z}$ with $\Delta_0 = 0$, $\Phi = 0.5\Phi_0$ and $k_z=0$. As the chemical potential $\mu$ increases, various $m_L$ are occupied. A topological transition can only occur for specific $m_L$ modes (highlighted in red). Consequently, $\mu$ must reach a threshold value $\mu^\text{ts}$, which in turn generates a trivial skin at the hybrid nanowire's end. (b) Same as (a) but for the second radial subband ($m_r = 1$). Notice that (a) [(b)] represents a case with unrealistically strong (realistically small) SOC. (c) Topological phase diagram as a function of $\mu$ and average SOC $\langle \alpha \rangle$ for the device studied in Fig.~\ref{Fig1}. Vertical lines mark the minimum $\mu$ required for the first possible topological transition in the $m_r = 0$ (purple) and $m_r = 1$ (blue) radial modes.}
   \label{Fig4}
\end{figure}

The above phenomenology is generic in full-shell nanowires, and ultimately leads to the trivial skin effect. To see this, it is sufficient to consider the normal-state of the uniform full-shell Hamiltonian by setting $\Delta_0 = 0$ and $\phi_g(z) = 0$ in Eq.~S5 of \SMref{A:full-shell}{S-II} and focusing on the electron block. In this case, the Hamiltonian commutes with the angular momentum $L_z = J_z - \frac{1}{2}n \tau_z = -i\partial_\varphi + \frac{1}{2}\sigma_z$, whose eigenvalues relate to $m_J$ through $m_J = m_L \oplus \frac{1}{2}n$. Consequently, only two of these angular modes ($m_L = \pm 1/2$ for $n=1$) corresponds to the required $m_J = 0$. Because the SOC is radial, the system develops Rashba-like subbands indexed by $m_L$ for each radial mode $m_r$, as depicted schematically in Fig.~\ref{Fig4}(a) for $m_r = 0$. The SOC energy shifts the energy minimum (the band bottom) away from the $m_J = 0$ modes that are capable of undergoing a topological transition (highlighted in red). All other modes below zero energy will be populated before the $m_J=0$ subband as we increase carrier density. We define $\mu^\text{ts}_{m_r = 0}$ as the chemical potential required to populate the $m_J=0$ modes.

To understand the implications of the chemical potential threshold during a tunneling spectroscopy experiment, we restore the $\phi_g(z)$ potential, responsible for depleting the nanowire at the interface with the tunnel barrier (i.e., at $z=0$). Moving inward from the interface, the local chemical potential $\mu(z)$ smoothly increases until it reaches $\mu_\text{bulk}$, as explained in \ref{A:phi(z)}. Because the topological transition demands that the chemical potential reaches a finite $\mu^\text{ts}_{m_r = 0}$, the transition cannot occur exactly at the interface. Instead, it occurs at a position $z$ deeper inside the wire where the local chemical potential reaches $\mu^\text{ts}_{m_r = 0}$. This spatial offset creates a topologically trivial superconducting section at the wire's end that we call the \emph{trivial skin}.

As shown in Fig.~\ref{Fig4}(b), populating the $m_J=0$ sectors in higher radial modes ($m_r>0$) requires even higher chemical potential thresholds ($\mu^\text{ts}_{m_r > 0}$) because the radial confinement energy adds to the SOC energy shift. This pushes the topological transition point even further into the wire. Figure~\ref{Fig4}(c) presents the topological phase diagram in the $ \mu $–$ \langle \alpha \rangle $ plane for the full-shell case studied in the main text. In this typical scenario, inducing a topological transition in the first radial mode ($m_r = 0$) would require an unrealistically strong SOC. Consequently, the experimentally relevant trivial skin arises from the combined effects of the SOC energy shift and the radial confinement energy. The resulting skin depth is correspondingly increased.

The phenomenology of a partial-shell nanowire stands in contrast. Partial-shell geometries lack rotational symmetry and do not exhibit fluxoid quantization. The topological mechanism is Zeeman-driven, and is consequently not restricted to $m_J=0$ modes: any subband can undergo a topological transition when subjected to an appropriate Zeeman field. As a result, the first filled subband becomes topological immediately above the depletion point. The lack of a finite chemical potential threshold ensures that the resulting MZM (or Q-MZM) is localized right at the interface between the tunnel barrier (insulator) and the hybrid nanowire (topological superconductor), eliminating any intervening trivial skin.

%------------------------------------------------------------
% APPENDIX: Parameters
%------------------------------------------------------------

\appendixpar{Parameters used in the figures}
\label{A:parameters}
We consider realistic Al/InAs hybrid nanowires. The effective electron mass is $m^* = 0.023 m_e$, where $m_e$ is the electron mass, the parent gap at zero magnetic field is $\Delta_0 = 0.23~\text{meV}$, the semiconductor-superconductor normal decay rate $\Gamma_\text{NS} = 3 \Delta_0$ and the Pauli paramagnetic limit is $V_C = 2 V_Z^\text{c}$. The partial-shell wire has bulk chemical potential $\mu_\text{bulk} = 2~\text{meV}$ and SOC $\alpha = 40~\text{meV}\,\text{nm}$. Figures~\ref{Fig1}(a–c) are calculated with smoothness parameter $\chi = 200~\text{nm}$. The full-shell wire has bulk chemical potential $\mu_\text{bulk} = 22.8~\text{meV}$, radial electrostatic potential energy variation $\Delta U = 60~\text{meV}$, average SOC $\langle \alpha \rangle = 7~\text{meV}\,\text{nm}$, superconductor-semiconductor normal decay rate $\Gamma_\text{NS} = 40 \Delta_0$, radius $R = 70~\text{nm}$, shell thickness $d = 10~\text{nm}$, shell diffusive coherence length $\xi_d = 70~\text{nm}$, and Landé factor $g = 10$. Figures~\ref{Fig1}(d–h) are calculated with smoothness parameter $\chi = 1~\mu\text{m}$. In Fig.~\ref{Fig3}, the disorder amplitude is chosen to be large, while ensuring that the MZM is not destroyed: for the partial-shell case [Figs.~\ref{Fig3}(a,b)] we take $A_\text{imp}/ \sqrt{N} = 0.32 \Delta_0$, whereas for the full-shell case [Figs.~\ref{Fig3}(c,d)] we use the smaller value $A_\text{imp} / \sqrt{N} = 0.02 \Delta_0$,  reflecting the smaller topological gap. $N$ is the number of sites where the disorder potential is applied.

\ifSMincluded
  \onecolumngrid
  \pagebreak
  \clearpage
  %\afterpage{%
    \setcounter{page}{1}%
    \renewcommand{\thepage}{S\arabic{page}}%
  %}%
  % counters that aren't page-shipout sensitive can be set normally:
  \setcounter{figure}{0}
  \setcounter{table}{0}
  \setcounter{equation}{0}
  \setcounter{section}{0}
  \renewcommand{\thetable}{S\arabic{table}}
  \renewcommand{\thefigure}{S\arabic{figure}}
  \setcounter{secnumdepth}{3}
  \renewcommand{\thesection}{S-\Roman{section}}
  \renewcommand{\thesubsection}{\Alph{subsection}}
  \begin{center}\textbf{\large Supplemental Material}\end{center}
  \vspace{0.5em}
  \FloatBarrier
  \twocolumngrid
  
%------------------------------------------------------------
% APPENDIX: Partial-shell
%------------------------------------------------------------

% Remove the asterisk (*) and the manual "S-I:" text
\section{Partial-shell nanowire model}
\label{A:partial-shell}
To capture the essential phenomenology of quasi-Majorana zero modes (Q-MZMs), we employ the standard single-channel Bogoliubov-de Gennes Hamiltonian \cite{Lutchyn:PRL10,Pan:PRR20},
\begin{equation}
    \label{eq:Hpartial}
    \begin{split}
    H_\text{PS} =& \left[\frac{p_z^2}{2 m^*} -\mu_\text{bulk} +e\phi_g(z)\right] \tau_z + V_Z \sigma_z +\\ & \alpha p_z \sigma_y \tau_z + \Sigma_\text{shell}(\omega, V_Z).
    \end{split}
\end{equation}
It is written in the Nambu basis $\left(\Psi_\uparrow, \Psi_\downarrow, \Psi_\downarrow^\dagger, -\Psi_\uparrow^\dagger\right)^T$ and $\sigma_i$($\tau_i$) are the Pauli matrices in spin (electron-hole) space. Setting $\hbar = 1$, the longitudinal momentum is $p_z = -i\partial_z$. The system parameters are defined as follows: $V_Z$ is the Zeeman energy ($V_Z=\mu_B gB/2$, where $\mu_B$ is the Bohr magneton, $g$ the Landé $g-$factor and $B$ the applied axial magnetic field), $\alpha$ is the Rashba SOC, $m^*$ is the effective electron mass, and $\mu_\text{bulk}$ is the chemical potential deep inside the wire. The electrostatic potential energy $e\phi_g(z)$ represents the semiconductor band bottom along the wire and has the profile defined in Appendix A of the main text, which ensures the necessary depletion at the tunnel barrier interface. Moreover, the superconductor degrees of freedom are integrated out, which introduces the self-energy,
\begin{equation}
    \Sigma_\text{shell}(\omega, V_Z) = - \Gamma_\text{NS} \sigma_0 \frac{\Delta(V_Z) \tau_x - \omega }{\sqrt{\Delta^2(V_Z) - \omega^2}},
\end{equation}
where the magnetic field-dependent superconducting gap is given by
\begin{equation}
\label{pauli}
    \Delta(V_Z) = \Delta_0 \sqrt{1 - \left(\frac{V_Z}{V_C}\right)^2}.
\end{equation}
Here, $\Delta_0$ is the parent gap, $V_C$ is the Pauli paramagnetic limit of the superconductor and $\Gamma_\text{NS}$ is the decay rate between the semiconductor and the shell in the normal state.

The topological phase diagram as a function of $\mu$ and $V_Z$ for the single-channel Hamiltonian is depicted in Fig. 1(c) and discussed in the main text. One may naturally consider multimode generalization of the single-channel model. When multiple transverse subbands are filled in a multimode nanowire, it will still become topological for $V_Z$ exceeding the critical Zeeman field of the shallowest populated subband. The topological phase diagram of Fig. 1(c) then develops additional topological regions, one per subband, at increasing values of $\mu$, with the overall topology following an even-odd pattern where these regions overlap \cite{Lutchyn:PRL11, Penaranda:PRR20}. In this work we do not include multimode effects for partial shell systems, however. The reason is that, in the presence of smooth confinement potentials, such multimode effects cannot give rise to the kind of trivial skin that appears in full-shell models. This stems from the fact that any partial-shell transverse mode can become topological when its corresponding critical Zeeman field is exceeded, while only $m_J=0$ modes can undergo a flux-driven transition in full-shell nanowires (see Appendix B of the main text for further discussion). Hence, a comparison to a simple, single-mode partial-shell model is sufficient when analyzing the implications of the trivial skin for MZM detection.

%------------------------------------------------------------
% APPENDIX: Full-shell
%------------------------------------------------------------

\section{Full-shell nanowire model}
\label{A:full-shell}
This appendix summarizes the key features of the full-shell nanowire model relevant to this work and closely follows Appendix A of Ref. \cite{Paya:PRB25}. The model is thoroughly described and discussed in Refs. \cite{San-Jose:PRB23, Paya:PRB24}. 

A full-shell nanowire consists of a semiconductor core proximitized on all facets by a diffusive superconducting shell. By integrating out the shell degrees of freedom, we obtain a self-energy $\Sigma_{\rm shell}$ acting on the core surface. The Green's function is then given by $G(\omega) = \left[\omega - H_{\rm core} - \Sigma_{\rm shell}(\omega) \right]^{-1}$, where $ H_{\rm core}$ is the Hamiltonian of the core. For analytical convenience, we express the model in terms of an effective Bogoliubov-de Gennes (BdG) Hamiltonian $H \equiv \omega - G^{-1}(\omega) = H_{\rm core} + \Sigma_{\rm shell}$, expressed in the Nambu basis $\Psi = \left(\psi_\uparrow, \psi_\downarrow, \psi^\dagger_\downarrow, -\psi^\dagger_\uparrow \right)$. 

Assuming a cylindrical symmetry for the full-shell nanowire [with coordinates $(r,\varphi,z)$], it can be shown that $\left[H, J_z\right] = 0$, where the generalized angular momentum $J_z = -i \partial_\varphi + \frac{1}{2}\sigma_z + \frac{1}{2}n \tau_z$ is the sum of the orbital angular momentum, spin momentum and ``fluxoid'' momentum. Here, $\sigma_i$ ($\tau_i$) are Pauli matrices for the spin (particle-hole) space, and $n$ is the fluxoid number. We can then block-diagonalize $H$ on subspaces with good quantum numbers $m_J$, the eigenvalues of $J_z$ \cite{Vaitiekenas:S20}, which take values
\beq
    m_J = \left\{\begin{array}{ll}
    \mathbb{Z}+ \frac{1}{2} & \textrm{if $n$ is even} \\
    \mathbb{Z} & \textrm{if $n$ is odd}
    \end{array}\right..
    \label{mJ}
\eeq
On $m_J$ subspaces, $H$ takes a $\varphi-$independent form,
\begin{equation}
    \label{eq:Hfull}
    \begin{split}
        H_\text{FS} =& \left[\frac{p_z^2 + p_r^2}{2 m^*} + U(r) + e\phi_g(r,z)- \mu_\text{bulk}\right]\sigma_0 \tau_z + V_Z \sigma_z \tau_0 \\
        &+ \frac{1}{2 m^* r^2}\left(m_J - \frac{1}{2}\sigma_z - \frac{1}{2}n\tau_z + \frac{1}{2}\frac{\Phi}{\Phi_0} \frac{r^2}{R_{\rm LP}^2}\tau_z\right)^2\sigma_0\tau_z \\
        &- \frac{\alpha(r)}{r}\left(m_J - \frac{1}{2}\sigma_z - \frac{1}{2}n\tau_z + \frac{1}{2}\frac{\Phi}{\Phi_0} \frac{r^2}{R_{\rm LP}^2}\tau_z\right)\sigma_z\tau_z \\
        & + \alpha(r)k_z\sigma_y\tau_z + \Sigma_{\rm shell}(\omega),
    \end{split}
\end{equation}
where $R_{\rm LP} = R + d/2$, $p_z = -i \partial_z$, $p_r = -i \partial_r$, $p_r^2 = -\frac{1}{r}\partial_r \left(r \partial_r\right)$, $m^*$ is the semiconductor effective mass, $\mu_\text{bulk}$ the chemical potential (with $\hbar = 1$) and $\Phi=\pi R_{\rm{LP}}^2B$ is the flux, where $B$ is the axial magnetic field. The Zeeman field is included as $V_Z = \frac{1}{2}g \mu_B B$, where $g$ is the Landé factor and $\mu_B$ the Bohr magneton. $\phi_g(r,z)$ is the electrostatic potential induced by the gates outside the wire, which decays as $z\rightarrow\infty$ (see Appendix A of the End Matter). $U(r)$ denotes the electrostatic potential energy at $z\rightarrow \infty$, deep inside the core. Although the precise form of $U(r)$ depends on the microscopic details of the interface, it is known that an epitaxial core/shell Ohmic contact leads to a dome-shaped profile \cite{Mikkelsen:PRX18, Antipov:PRX18}, which we model as
\begin{equation}
    U(r) = \Delta U \left[1-\left(\frac{r}{R}\right)^2\right].
\end{equation}
We assume a Rashba type SOC arising from the inversion symmetry breaking at the core/shell interface, which is radial and points outwards, $\alpha(r) = -\alpha_0 \partial_r U(r)$, with $\alpha_0$ a model parameter.

Radial confinement leads to quantized radial modes labeled with the integer quantum number $m_r \geq 0$, which counts the number of radial noes of the wavefunction. However, the Majorana condition, $\mathcal{P} \Psi_{m_J}(\omega = 0) = \Psi_{-m_J}(\omega = 0) = \Psi_{m_J}(\omega = 0)$, can only be satisfied in the $m_J = 0$ sector. Each radial mode with $m_J = 0$ undergoes a separate topological phase transition at distinct points of parameter space, as depicted in Fig. 1(e, f). Notice that the topological region associated with $m_r = 0$ occurs at values of $\langle \alpha \rangle$ outside the range explored in this work.

The self-energy of the tubular superconducting shell needs to incorporate pair-breaking effects due to the magnetic flux. These effects are stronger than the pair-breaking from the Zeeman effects, Eq. \eqref{pauli}, that dominate the partial-shell case, so that the latter is neglected. Orbital pair-breaking is included instead within a diffusive superconductor approximation. Following Ref. \cite{Skalski:PR64}, this results in a self energy
\begin{equation}
    \Sigma_{\rm shell}(\omega) = \Gamma_\text{NS} \sigma_0 \frac{\tau_x - u(\omega)\tau_0}{\sqrt{1 - u(\omega)^2}},
\end{equation}
where $u(\omega)$ satisfies
\begin{equation}
    u(\omega) = \frac{\omega}{\Delta (\Lambda)} + \frac{\Lambda}{\Delta (\Lambda)} \frac{u(\omega)}{\sqrt{1-u(\omega)^2}},
\end{equation}
and  $\Lambda$ is a depairing parameter. The superconducting pairing amplitude $\Delta$ obeys
\begin{equation}
    \begin{split}
        \ln\frac{\Delta(\Lambda)}{\Delta(0)} =& -P\left(\frac{\Lambda}{\Delta(\Lambda)}\right),\nonumber\\
P(\lambda\leq 1) =&\frac{\pi}{4}\lambda,\nonumber\\
P(\lambda\geq 1) =& \ln\left(\lambda+\sqrt{\lambda^2-1}\right)+\frac{\lambda}{2}\arctan\frac{1}{\sqrt{\lambda^2-1}}\nonumber\\
&-\frac{\sqrt{\lambda^2-1}}{2\lambda}.
    \end{split}
\end{equation}
We select the solution for $u(\omega)$ that ensures the correct continuity and asymptotic behavior of the retarded Green's function. The superconducting gap is given by \cite{Skalski:PR64}
\begin{equation}
    \Omega(\Lambda) = \left(\Delta(\Lambda)^{2/3} - \Lambda^{2/3}\right)^{3/2}.
\end{equation}
Note that, at zero depairing (or, equivalently, at $B=0$), $\Omega(0)=\Delta(0)\equiv\Delta_0$, the parent superconductor gap.
The relation between the depairing parameter and the magnetic flux is \cite{Schwiete:PRB10}
\begin{equation}
   \begin{split}
       \Lambda(\Phi) &= \frac{k_{B} T_{\rm c}\,\xi_{d}^2}{\pi R_{\rm{LP}}^2}\left[4\left(n-\frac{\Phi}{\Phi_0}\right)^2 + \frac{d^2}{R_{\rm{LP}}^2}\left(\frac{\Phi^2}{\Phi_0^2} + \frac{n^2}{3}\right)\right],\\
    n(\Phi) &= \lfloor \Phi/\Phi_0\rceil = 0, \pm 1,\pm 2, \dots,
   \end{split}
   \label{depairing}
\end{equation}
where $\xi_d$ is the diffusive superconducting coherence length, $T_c$ is the zero-flux critical temperature and $k_B$ the Boltzmann constant.

%------------------------------------------------------------
% APPENDIX: Numerics
%------------------------------------------------------------

\section{Numerical implementation and observables}
\label{A:numerics}
Numerical calculations, described in detail in Refs. \cite{San-Jose:25a, Paya:26a, Paya:26}, rely on a tight-binding description of the effective Hamiltonians, with discretization lattice parameter $a_0 = 5$~nm. Topological phase diagrams are computed through the topological invariant at $k_z = 0$ (and $m_J = 0$ for the full-shell model) (see Ref. \cite{Paya:PRB24}),
\begin{equation}
    \mathcal{Q} = \text{sign}\left\{\text{Pf}\left[\sigma_y \tau_y H\left(k_z = 0\right) \right]\right\}.
\end{equation}
The LDOS is computed through the retarded Green's function $G^r = \left(\omega + i0^+ - H\right)^{-1}$,
\begin{equation}
    \text{LDOS} = -\frac{1}{\pi} \text{Im}\left[\text{Tr}\left(G^r\right)\right],
\end{equation}
while the DOS is defined as the integral of the LDOS along the wire axis up to the point where $\mu(z) \sim \mu_\text{bulk}$.

The differential conductance is calculated by integrating out the normal lead to obtain a self-energy $\Sigma_\text{lead}$ acting on the nanowire. For simplicity, the lead is modeled as identical to the bulk hybrid nanowire (setting $\Gamma_\text{NS} = 0$). The lead-nanowire coupling is incorporated within the tight-binding formalism as a fraction of the intra-wire hopping. The zero-temperature differential conductance is then
\begin{equation}
    \label{eq:conductance}
    \frac{dI}{dV}(V) = \frac{e^2}{h} \text{Tr}\left[i \left(G^r - G^a\right)\Gamma \tau_e - G^r \Gamma \tau_zG^a\Gamma\tau_e\right],
\end{equation}
where $G^{r(a)}$ is the retarded (advanced) Green's function of the hybrid nanowire, and $\Gamma = i \left(\Sigma_\text{lead} - \Sigma_\text{lead}^\dagger\right)$. Here, $\tau_j$ are the Pauli matrices in Nambu space, and $\tau_e = \frac{1}{2}\left(\tau_0 + \tau_z\right)$. The finite-temperature conductance is obtained by convolving Eq. \eqref{eq:conductance} with the thermal broadening kernel,
\begin{equation}
    \label{eq:conductanceT}
    \frac{dI}{dV}(V, T) = \int d\omega \left[\frac{\partial \left(\omega - eV, T\right)}{\partial \omega}\right] \frac{dI}{dV}(\omega).
\end{equation}
%------------------------------------------------------------
% APPENDIX: Conductance
%------------------------------------------------------------

\section{Differential conductance simulations}
\label{A:conductance}

\begin{figure}
   \centering
   \includegraphics[width=\columnwidth]{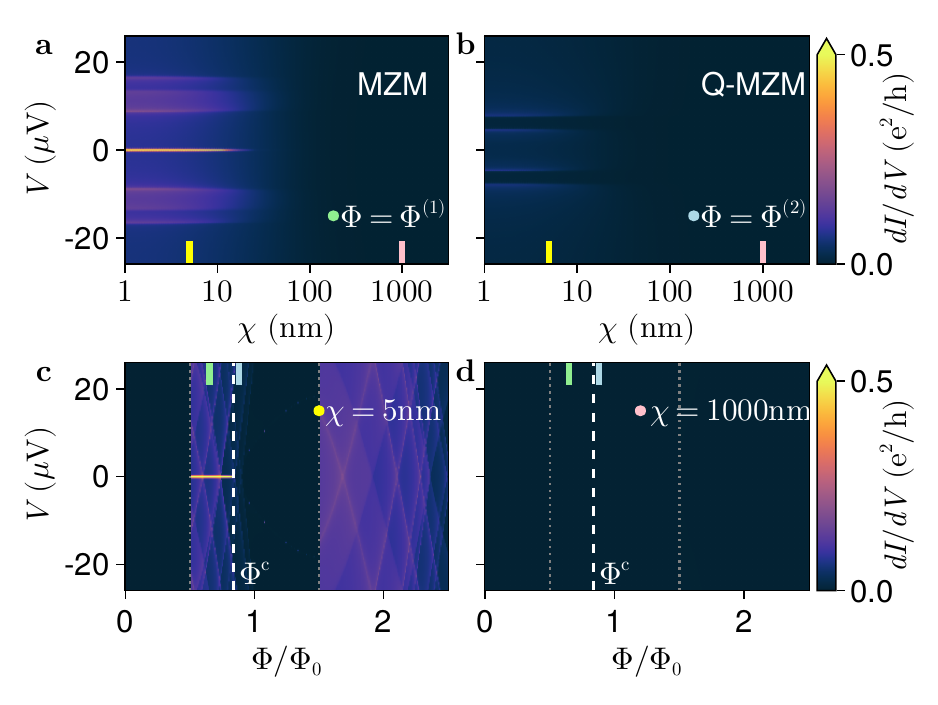}
   \caption{\textbf{MZMs and Q-MZMs probed through tunneling spectroscopy.} (a, b) Differential conductance $dI/dV$ through a tunnel barrier at the end of the semi-infinite full-shell nanowire studied in Fig. 1 and Fig. 2(c,d) of the main text, as a function of bias voltage $V$ and smoothness parameter $\chi$, including all angular modes at very low temperature $T=1$mK. Panel (a) [(b)] is calculated at flux $\Phi^{(1)}$ ($\Phi^{(2)}$), corresponding to the MZM (Q-MZM) case. (c, d) $dI/dV$ a function of $V$ and normalized flux $\Phi/\Phi_0$, for the same device, comparing a steep [$\chi = 5$~nm, panel (c)] and a smooth [$\chi = 1000$~nm, panel (d)] confining potential.}
   \label{fig:conductance}
\end{figure}

% Panel (a) is calculated at a flux $\Phi = \Phi^{(1)}$ where the entire hybrid nanowire is in the topological regime. A zero bias peak (ZBP) is present because the trivial skin region is sufficiently short to keep the MZM close to the end of the wire. In panel (b), calculated at flux $\Phi = \Phi^{(2)}$, only a section of the wire is in the topological regime. Andreev bound states (ABSs) that coalesce into a Q-MZM are visible only at low $\chi$, whereas the trivial skin hides the Q-MZM ZBP at higher smoothness.
% A ZBP is  visible only within the topological regime ($\Phi < \Phi^c$)  in the steep case.  Outside the topological regime, the trivial skin effect prevents the appearance of a ZBP. Throughout all panels, CdGM analogs are present; while they fill the topological gap, they do not interact with the MZM.

Here, we expand upon the results of Fig. 2 in the main text by simulating the local differential conductance. We model the hybrid nanowire coupled to a normal lead via a tunnel barrier, calculating the differential conductance $dI/dV$ as a function of the applied bias voltage $V$. We consider the contribution of not only the $m_J=0$ sector, but also all higher-order angular modes ($m_J \neq 0$). While these modes cannot undergo a topological phase transition, they host Caroli-de Gennes-Matricon (CdGM) analog states. 

In Fig. \ref{fig:conductance}(a, b) we show $dI/dV$ simulations versus $V$ and smoothness parameter $\chi$ for two representative fluxes. For $\Phi = \Phi^{(1)}$ [panel (a)], the bulk of the hybrid nanowire is in the topological regime, localizing a Majorana zero mode (MZM) at the end of the trivial skin region (see Fig. 1(d1) of the main text). Consistent with the LDOS calculation of Fig. 2(c) of the main text, this MZM manifests as a robust zero-bias peak (ZBP) for sufficiently steep confinement (small $\chi$). As $\chi$ increases, the trivial skin widens and pushes the MZM deeper into the hybrid nanowire, exponentially suppressing its visibility in tunneling spectroscopy at finite temperatures \footnote{Technically, the $T\to 0$ limit is singular since, at $T=0$, even a buried MZM or Q-MZM with exponentially small coupling to the lead would be visible as a $dI/dV\sim 2e^2/h$ ZBP, albeit of exponentially small width versus bias. However, for any small but finite temperature greater that said width ($1$mK in this section), the peak visibility becomes exponentially suppressed, and $dI/dV$ becomes a faithful measurement of the LDOS.}. At small $\chi$'s, an additional differential conductance is present for all (subgap) $V$ values, corresponding to the CdGM analogs associated with the $m_J \neq 0$ modes (as well as $m_J=0$. The CdGM analog tails contribute with a small (purple) conductance background, whereas the Van-Hove singularities are seen with a brighter color at finite bias. Since these states are mostly delocalized along the hybrid wire, they contribute with a smaller $dI/dV$ than the MZM, which is a bound state close to the nanowire's end. Since CdGM analogs are uncoupled from the MZM, they do not obscure the MZM ZBP, provided that the Van-Hove peaks do not accidentally cross zero energy at a specific applied flux. Again, as $\chi$ increases, the contribution of the CdGM analogs to the $dI/dV$ also fades away.
%Because these states are not strictly localized at the boundary, they only appear where their parent angular subband is occupied; thus, their tunneling signatures also fade at large $\chi$.
%Although these van Hove singularities populate the topological gap, they remain uncoupled from the MZM. Consequently, they do not obscure the MZM ZBP, provided their divergent peaks do not accidentally cross zero energy at a specific applied flux.

At $\Phi = \Phi^{(2)}$ [panel (b)], only a finite section of the hybrid nanowire satisfies the topological conditions, while the bulk remains trivial (see Fig. 1(d2) of the main text). Again, consistent with Fig. 2(d) of the main text, for sufficiently small $\chi$ values, no Q-MZMs are possible and the $dI/dV$ contains only signatures of split MZMs and CdGM analogs. When the Q-MZM regime is reached for larger $\chi$ values, the size of the trivial skin is too large to allow for detection of the associated ZBP.

%As happened with the anal LDOS calculations of the main text, when the Q-MZM regime is reached for large $\chi$, the size of the trivial skin is too large to allow for the detection of the associated ZBP. Therefore, the only signature of this topological region are split MZMs, or ABSs, at low $\chi$, indistinguishable from any other CdGM analog.

In actual experiments, the smoothness parameter $\chi$ is not really tunable. For this reason, in Fig. \ref{fig:conductance}(c, d) we show $dI/dV$ simulations versus $V$ and $\Phi/\Phi_0$ for two representative values of $\chi$. As discussed in the main text, MZMs can only appear in odd Little-Parks lobes and up to a critical flux value $\Phi^c$. For a hard confinement (small $\chi$), Fig. \ref{fig:conductance}(c), a ZBP associated to a MZM is visible within the topological flux window over a small background of CdGM analogs that fill the topological gap but do not interact with the MZM. The lobe $0$ and the rest of  lobe $1$ are empty for the voltage-bias range shown, while the lobe $2$ is filled with the CdGM analog contribution. For a smooth confinement (large $\chi$), Fig. \ref{fig:conductance}(d), MZMs and Q-MZMs localize deep into the wire after the trivial skin, so they are not visible through local conductance. The same applies for the CdGM analog contribution.

In summary, differential conductance simulations (including all $m_J$ sectors) confirm the conclusions reached in the main text through $m_J=0$-LDOS calculations. Q-MZM ZBPs in tunneling spectroscopy are effectively invisible in full-shell hybrid nanowires (in contrast to what happens in partial-shell geometries), due to the emergence of a trivial skin that effectively shields Q-MZMs from detection.

%------------------------------------------------------------
% APPENDIX: Temperature
%------------------------------------------------------------

\section{Finite-temperature simulations}
\label{A:temperature}

\begin{figure}
   \centering
   \includegraphics[width=\columnwidth]{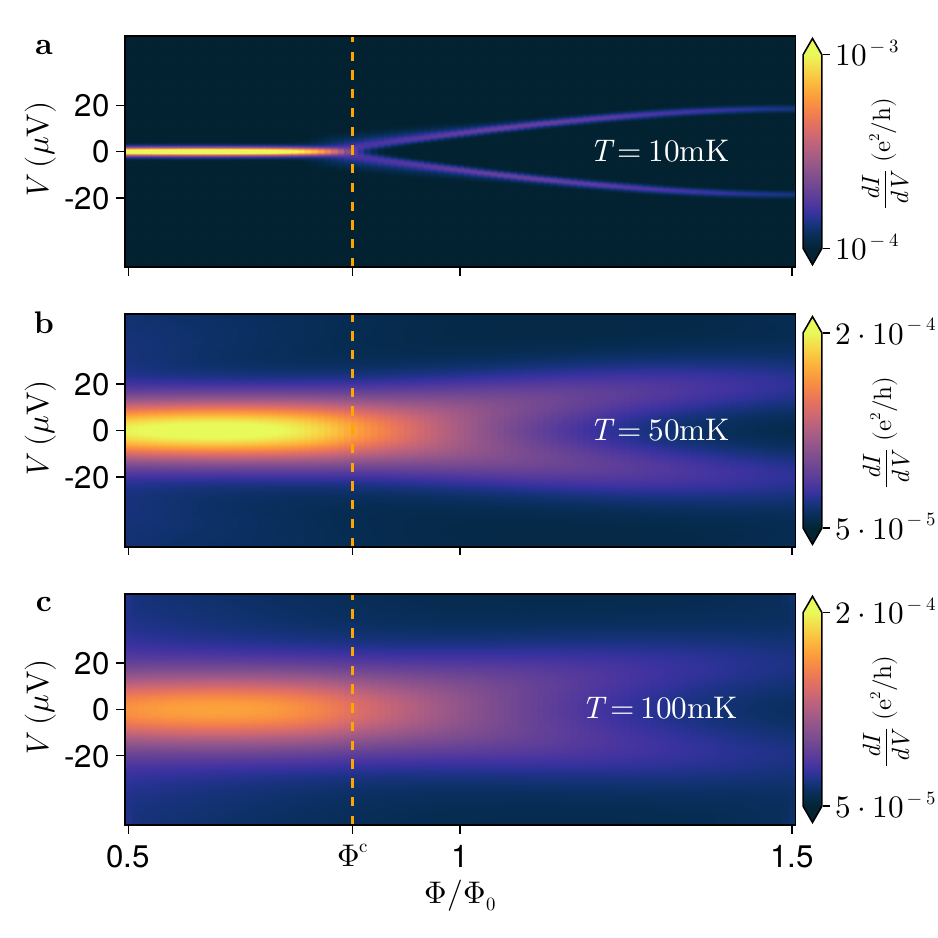}
    \caption{\textbf{Temperature dependence of tunneling differential conductance.} Contribution of the  $m_J = 0$ angular mode to the differential conductance at $\chi = 10$nm for increasing $\Phi$ in the first LP lobe and temperatures $T=10, 50, 100$mK. The dashed line marks the topological transition $\Phi_c$ separating a MZM (left) and two split Q-MZMs (right).}
    \label{fig:temperature}
\end{figure}
%For low temperatures (a, b), the MZM zero bias peak (ZBP) is clearly distinguishable from two closely spaced ABSs. For higher temperatures (c, d), the peaks broaden until an MZM ZBP cannot be distinguished from two closely spaced ABS peaks (e, f).

In \ref{A:conductance}, we provided tunneling conductance simulations at very low temperatures ($T=1$mK) to show that, at any small but non-zero temperature, the conclusions derived from the LDOS in the main text also extend to $dI/dV$, which is a faithful proxy of the LDOS. In a physical experiment, however, conductance peaks are inevitably broadened by thermal fluctuations. This can pose a problem to experimentally determine whether a single $dI/dV$ peak around zero bias is an actual MZM of bulk-topological origin or rather a broadened double-peak from two near-zero modes, as those arising at small $\chi$ from the splitting of two Q-MZMs. Such a situation has the potential to introduce temperature-induced false positives where weakly split Q-MZMs are mistaken with a MZM. 

Interestingly, full-shell nanowires provide a natural way to identify temperature-induced false positives. The key is already apparent in Fig. 1(h) of the main text. The splitting of a pair of Q-MZMs at small $\chi$ quickly grows with flux $\Phi$, while the width of a MZM ZBP is, in contrast, $\Phi$-independent. Hence, a thermally-broadened double-peak in $dI/dV$ will naturally resolve into two distinct peaks as $\Phi$ is increased. This is shown in Fig. \ref{fig:temperature}, where we plot the full-shell $dI/dV$ at small $\chi = 5$nm for different temperatures as $\Phi$ is increased across the topological transition (dashed lines). The $dI/dV$ exhibits a MZM on the left and split Q-MZMs on the right, with an increasing thermal broadening in each panel. To directly mirror the results in Fig.~2(c,d) of the main text, we restrict this analysis to the $m_J=0$ sector. We see that, indeed, the merged double-peak in the trivial region eventually bifurcates into two distinct peaks upon increasing $\Phi$.
We thus conclude that temperature-induced false positives at small $\chi$, which may arise when the splitting of two Q-MZMs is smaller than temperature, can be readily identified by analyzing peak behavior with flux.

% It therefore becomes important to determine if typical Q-MZM splittings expected in full-shell nanowires are greater than relevant temperatures in the lab. If they are not, temperature would become a source of a different type of false positive in the presence of small-$\chi$ split Q-MZMs.

% Figure \ref{fig:temperature} shows the calculated tunneling differential conductance at various temperatures as a function of bias voltage $V$ and the smoothness parameter $\chi$. To directly mirror the results in Fig.~2(c,d) of the main text, we restrict this analysis to the $m_J=0$ sector. In the left column, the applied flux places the bulk hybrid nanowire in the topological regime, revealing a thermally broadened MZM peak at small values of $\chi$. In the right column, the system is tuned to the Q-MZM regime. Here, at small $\chi$, two thermally broadened ABSs are visible, which merge into a single peak as temperature increases. We find that at typical temperatures below $\sim 25$mK, it is possible to distinguish MZMs from small-$\chi$ split Q-MZMs. While the value of this threshold temperature may depend on parameters of the full-shell wire, we estimate this to be the typical threshold  below which temperature-induced false positives can be excluded.
%Ultimately, resolving a true Majorana ZBP requires an operating temperature low enough that the thermal broadening remains smaller than the intrinsic energy splitting of the adjacent subgap states.
\fi

\end{document}